\documentclass[aps,prd,amsmath,amssymb,showpacs]{revtex4}

\usepackage{graphicx}
\begin{document}

\title{Realistic constraints on the doubly charged bilepton 
couplings from Bhabha scattering with LEP data}
\author{S. Ata\u{g} }
\email[]{atag@science.ankara.edu.tr}
\affiliation{Department of Physics, Faculty of Sciences, 
Ankara University, 06100 Tandogan, Ankara, Turkey}

\author{K.O. Ozansoy }
\email[]{oozansoy@science.ankara.edu.tr}
\affiliation{Department of Physics, Faculty of Sciences,
Ankara University, 06100 Tandogan, Ankara, Turkey}

\begin{abstract}
Upper limits on doubly charged bilepton couplings and masses 
are extracted from LEP data for Bhabha scattering at energy 
range $\sqrt{s}=183-202$ GeV using standard model program ZFITTER
which calculates radiative corrections. We find that 
$g_{L}^{2}/M_{L}^{2}<O(10^{-5})GeV^{-2}$ at 95\% C.L. 
for scalar and vector bileptons.
\end{abstract}


\maketitle

\section{Introduction}
The success of the standard model of strong and electroweak 
interactions has been verified experimentally with high accuracy 
at present collider energies. Based on the common opinion
that the standard model is not the final theory of elementary
particle world,  many extensions of the 
standard model are studied extensively to look for new physics 
beyond it at higher energies. 

New physics has been searched via several exotic particles 
such as leptoquarks, bileptons, diquarks and many more 
which are not covered by the standard model. In this work 
we study bileptons which are defined as bosons carrying lepton 
number L=2 or 0. Therefore they couple to two standard model 
leptons but not to quarks. Bileptons appear in models where 
$\text{SU(2)}_{\text{L}}$ gauge group is extended 
to SU(3) \cite{frampton}. Bileptons are 
also obtained in models with extended Higgs sectors that contain 
doubly charged Higgs bosons \cite{haber}. Grand unified theories, 
technicolor and 
composite models predict the existence of bileptons as well as 
other exotic particles \cite{ross}. Classification and interactions of 
bileptons are provided by several works \cite{rizzo} and  
a comprehensive review has been presented  in \cite{cuypers}
including low and high energy bounds on their couplings.
Indirect constraints on the masses and couplings of doubly 
charged bileptons have been obtained from lepton number 
violating processes and muonium-antimuonium conversion 
experiments \cite{swartz,sasaki,chang,willmann}. 

A recent search for doubly charged bilepton (as Higgs boson) 
has been performed by DELPHI collaboration at LEP2
\cite{delphi}

At LEP fermion pair production is the unique 
reaction to test the standard model at loop
correction level \cite{sirlin}. 
Therefore one needs precision calculations 
including QED and weak corrections for reliable comparison 
with experiments. ZFITTER is one of the standard model 
program developed to compute scattering cross sections 
and assymmetries for fermion pair production 
in $\text{e}^{+}\text{e}^{-}$ 
collision with QED and electroweak corrections
\cite{zfitter}. Using cross section
calculations with ZFITTER realistic limits for new physics 
can be obtained from LEP data. 
Since QED corrections are model independent (well-defined 
if couplings, mass and width of new particles are fixed), the usual 
convolution formulae can be applied for total cross section 

\begin{eqnarray}
\sigma(s)=\int{dv~ \sigma^{0}(s^{\prime})~R(v)} 
\end{eqnarray}
with $v=1-s^{\prime}/s$ and the flux factor $R(v)$ 
(radiator) is not influnced by new particles such as bileptons.
Above equation can be straightforwardly generalized to 
different assymmetries $A_{FB}$, $A_{pol}$, $A_{LR}$ or to 
scattering angle distribution $d\sigma/\cos{\theta}$ 
with different effective Born terms and radiators. Final state 
acollinearity cut and minimum energy can also be applied.

Contribution of doubly charged bileptons to Bhabha scattering 
is the subject of the present article  

\begin{eqnarray}
\text{e}^{+}\text{e}^{-}\to (\gamma, Z, L^{--})\to 
\text{e}^{+}\text{e}^{-} (\gamma)
\end{eqnarray}
where $L^{--}$ is doubly charged bilepton.
With the  influence of bileptons  the cross section covers: 
standard model terms with electroweak and QED corrections 
arising from $\gamma$, $Z$ exchanges 
and their interferences;
virtual bilepton exchange with QED correction as explained above; 
interference terms of bilepton with $\gamma$, $Z$.

For complete radiative corrections to Bhabha scattering ZFITTER has 
limitation especially for higher acollinearity angles.
But for small acollinearity angles the results are reasonable for our 
purpose.

\section{Interaction lagrangian and Bhabha scattering}
General effective lagrangian describing interactions of 
bileptons with the standard model leptons is generated by 
requiring $\text{SU(2)}_{\text{L}}\times \text{U(1)}_
{\text{Y}}$ invariance. We consider lepton flavor conserving 
part of the lagrangian and L=2 bileptons as follows: 

\begin{eqnarray}
{\cal L}_{L=2}=&&g_1 \bar{\ell}^{c}i\sigma_2 \ell L_1+\tilde{g}_1
\bar{e}_{R}^{c}e_{R}\tilde{L}_1 \nonumber \\
+&&g_2\bar{\ell}^{c}i\sigma_2\gamma_{\mu}
e_{R}L_{2}^{\mu}+g_3\bar{\ell}^{c}i\sigma_2\vec{\sigma}
\ell~.\vec{L}_3+h.c.
\end{eqnarray}

In the notations we have used ${\ell}$ is the left handed 
$\text{SU(2)}_\text{L}$ lepton doublet and $e_{R}$ is the 
right handed charged singlet lepton. Charge conjugate fields are 
defined as $\psi^{c}=C\bar{\psi}^T$ and $\sigma_1$, 
$\sigma_2$, $\sigma_3$ are the Pauli matrices.  
The subscript of bilepton fields $L_{1,2,3}$ and 
couplings $g_{1,2,3}$ denote $\text{SU(2)}_\text{L}$ singlets, 
doublets and triplets.

Here we are interested only in doubly charged bileptons. 
In order to express the lagrangian in terms of individual electron, 
bileptons and helicity projection operators 
$\text{P}_{\text{R/L}}=\frac{1}{2}(1\pm\gamma_5)$
we expand the Pauli matrices and lepton doublets
and write the lagrangian as :
\begin{eqnarray}
{\cal L}_{L=2}=&&\tilde{g}_1~\tilde{L}_1^{++}~\bar{e}^{c}P_R e
+g_2~ L_{2\mu}^{++}~\bar{e}^{c}\gamma^{\mu}P_L e
\nonumber\\
&&-\sqrt{2}g_3~ L_3^{++}~\bar{e}^{c}P_L e + h.c.
\end{eqnarray}
where superscripts of bileptons stand for their electric charges.

Doubly charged bileptons contribute Bhabha scattering through 
their u-channel exchange. For the scalar $\text{L}_{3}^{--}$ 
exchange, the polarized differential cross section in terms 
of mandelstam invariants s, t and u is given by

\begin{eqnarray}
\frac{d\sigma}{d\cos{\theta}}=&&\frac{\pi\alpha^{2}}{2s}
\left\{ (\lambda_1+\lambda_2)\left[ 2(\frac{u}{t}+\frac{u}{s})
\right. \right.  \notag \\
&&\left. +2C_L^{2}(\frac{u}{t-M_{Z}^2}+\frac{u}{s-M_{Z}^2})+
\frac{g_L^2}{g_e^2}\frac{u}{u-M_{L}^2}\right]^2  \notag \\
&&+(\lambda_1-\lambda_2)\left[2(\frac{u}{t}+\frac{u}{s})+
2C_R^{2}(\frac{u}{t-M_{Z}^2}+\frac{u}{s-M_{Z}^2})\right]^2 
 \notag \\
&&+2\lambda_1\left[\frac{2t}{s}+C_L C_R\frac{2t}{s-M_{Z}^2}
\right]^2  \notag \\
&&\left. + 2\lambda_3\left[\frac{2s}{t}+C_L C_R\frac{2s}{t-M_{Z}^2}
\right]^2 \right\}   
\end{eqnarray}
where the definition of mandelstam variables, polarization factors and 
$C_L$,  $C_R$ are as follows:
\begin{eqnarray}
&&t=-\frac{s}{2}(1-\cos{\theta}), ~~
u=-\frac{s}{2}(1+\cos{\theta}),  \\
&&C_L=\frac{2\sin^2{\theta_W}-1}{2\sin{\theta_W}\cos{\theta_W}}, 
~~C_R=\tan{\theta_W},\\
&&\lambda_1=1-P_{-}P_{+}, ~~\lambda_2=P_{+}-P_{-}, \nonumber \\
&& \lambda_3=1+P_{-}P_{+}
\end{eqnarray}
Bilepton coupling $\sqrt{2}g_{3}$ in the lagrangian 
has been replaced by $g_{L}$ in the cross section and  $P_{\pm}$ is 
the polarization of positron or electron beams. Electromagnetic 
coupling $g_e$ is defined by $g_{e}^{2}=4\pi\alpha$. Together with 
polarization  factors the differential cross section is consistent 
with the one used in  ZFITTER  in the case of the standard model.

For vector bilepton $L_{2\mu}^{--}$ exchange with replacement 
$g_{2}\to g_{L}$ 
the differential cross section can be found as below:

\begin{eqnarray}
\frac{d\sigma}{d\cos{\theta}}=&&\frac{\pi\alpha^{2}}{2s}
\left\{ (\lambda_1+\lambda_2)\left[ 2(\frac{u}{t}+\frac{u}{s})
\right. \right.  \notag \\
&&\left. +2C_L^{2}(\frac{u}{t-M_{Z}^2}+\frac{u}{s-M_{Z}^2})
\right]^2  \notag \\
&&+(\lambda_1-\lambda_2)\left[2(\frac{u}{t}+\frac{u}{s})+
2C_R^{2}(\frac{u}{t-M_{Z}^2}+\frac{u}{s-M_{Z}^2})\right]^2
 \notag \\
&&+(\lambda_1+\lambda_2)\left[\frac{2t}{s}
+C_L C_R\frac{2t}{s-M_{Z}^2}
-\frac{g_L^2}{g_e^2}\frac{2t}{u-M_{L}^2}  \right]^2  \notag \\
&&+(\lambda_1-\lambda_2)\left[\frac{2t}{s}
+C_L C_R\frac{2t}{s-M_{Z}^2} \right]^2 \notag \\
&&\left. + 2\lambda_3\left[\frac{2s}{t}+C_L C_R\frac{2s}{t-M_{Z}^2}
\right]^2 \right\}
\end{eqnarray}

Investigation of new physics beyond the standard model can be
divided into three regions according to new physics energy
scale $\Lambda$ : $\sqrt{s}<< \Lambda$ leads to effective
lagrangian, four fermion contact interactions;
$\sqrt{s}>> \Lambda$ new physics, new particles;
$\sqrt{s}\sim \Lambda$ strongly depends on model.
At LEP energies we assume that bilepton masses are larger 
than $\sqrt{s}$ and $u<<M_{L}^{2}\sim\Lambda^{2}$ . Then 
we take into account the approximation
in the bilepton propagator:

\begin{eqnarray}
\frac{g_{L}^2}{u-M_{L}^2}\simeq\frac{g_{L}^2}{M_{L}^2}
\end{eqnarray}
which reduces the number of parameters.

\section{Results and Discussion}

Based on the previous considerations, bilepton parts of the 
cross section have been included into BHANG which runs 
together with ZFITTER6.36. 

\begin{figure}
\includegraphics{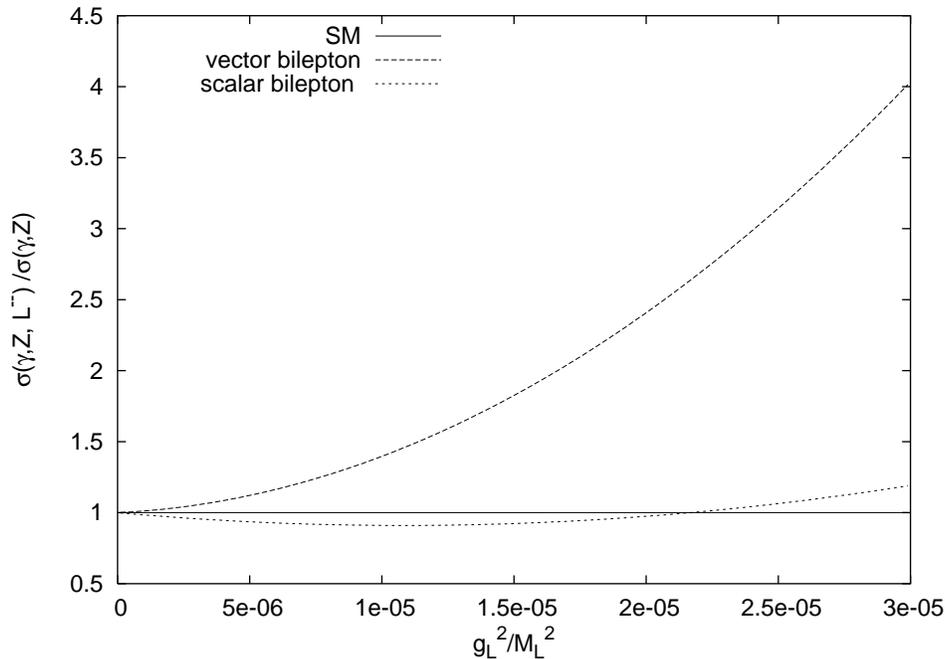}
\caption{Ratios of cross sections 
$\sigma(\gamma, Z, L^{--})/\sigma(\gamma, Z)$
from ZFITTER with and without bilepton exchange. 
\label{fig1}}
\end{figure}

Ratios of Bhabha scattering cross sections from ZFITTER 
with and without bilepton exchange 
$\sigma(\gamma, Z, L^{--})/\sigma(\gamma, Z)$
are shown in Fig.~\ref{fig1} as a function of $g_{L}^2/M_{L}^2$  
at $\sqrt{s}=192$ GeV for scalar and vector bileptons.
Curve corresponding vector bilepton exchange deviates 
more rapidly from standard model curve than the 
case of scalar one. 

Measured cross sections with OPAL detector at LEP have been used 
for our analysis \cite{opal1}. In order to 
give an idea about the comparison 
Table~\ref{tab1} shows the experimental values of the total 
cross sections and predicted standard model values from ZFITTER6.36
at LEP2 energy region $\sqrt{s}=183-202$ GeV for two 
scattering angular regions. Since ZFITTER is not good for 
high acollinearity angles in the case of Bhabha scattering 
we have chosen the measured cross sections only for 
$\theta_{acol}<10^{o}$.

We have used simple $\chi^2$ criterion from measured cross section 
to estimate upper limit on $g_{L}^2/M_{L}^2$. For combined results 
following expression can be considered:

\begin{eqnarray}
\chi^{2}=\sum_{i}(\frac{\sigma_{i}^{exp}-\sigma_{i}^{new}}
{\Delta_{i}^{exp}})^{2} 
\end{eqnarray}

\begin{eqnarray}
\Delta^{exp}=\sigma^{exp}\sqrt{\delta_{stat}^{2}+
\delta_{sys}^{2}}
\end{eqnarray}
where $i$ represents energy index corresponding energy values, 
cross sections and experimental errors in Table~\ref{tab1}.
Using the equation $\chi^{2}-\chi_{min}^{2}=2.7$
for one parameter, one sided analysis at 95\% 
confidence level, the upper limits on the 
$g_{L}^2/M_{L}^2$ for scalar and vector
bileptons are provided in 
Table~\ref{tab2} and Table~\ref{tab3}.
From tables, combined results for upper limits 
are $3.5\times 10^{-5},~~1.9\times 10^{-4}$ in the case of 
scalar bileptons and $9.8\times 10^{-6},~~7.4\times 10^{-5}$ 
for vector bileptons depending on angular regions.

Present limits for doubly charged bileptons are given below 
to compare with our results: 
$g_{ee}g_{\mu\mu}/M^{2}<5.8\times 10^{-5}$ at 90\% C.L. 
(from muonium-antimuonium conversion), 
$g_{ee}^{2}/M^{2}<9.7\times 10^{-6}$
at 95\% C.L. (from Bhabha scattering)
in Ref.~\onlinecite{swartz}; 
$g_{3}/M<(2600)^{-1}$ from muonium-antimuonium conversion,
at 95\% C.L. in Ref.~\onlinecite{willmann};
$g_{3}/M<(215)^{-1}$
at 90\% C.L. in Ref.~\onlinecite{sasaki}; $g_{3}/M<(340)^{-1}$
at 95\% C.L. in Ref.~\onlinecite{frampton2}. Here all masses are 
in GeV units.

In conclusion, it is probable that if complete radiative 
corrections to Bhabha scattering are realized
by ZFITTER as well as $e^{+}e^{-}\to \mu^{+}\mu^{-}$ process, 
more stringent and realistic  limits will be obtained for 
doubly charged bileptons.

\begin{table}
\caption{Measured cross sections with OPAL detector 
at LEP for two different angular regions 
with $\theta_{acol}<10^{o}$ . The  first error shown 
is statistical, the second systematic. The standard model 
cross sections are predicted by ZFITTER6.36\label{tab1}}
\begin{ruledtabular}
\begin{tabular}{llrlr}
$\sqrt{s}$ GeV & $\sigma$(pb) & $\sigma^{SM}$(pb) &$\sigma$(pb) &
$\sigma^{SM}$(pb)\\
 &$|\cos\theta|< 0.7$ &$|\cos\theta|< 0.7$ &$|\cos\theta|< 0.96$ &
 $|\cos\theta|< 0.96$\\
\hline
183 & 21.7$\pm$0.6$\pm$0.2 &20.9 &333.0$\pm$3.0$\pm$4.0 &320.7 \\
189 &20.08$\pm$0.33$\pm$0.10 & 19.55 &304.6$\pm$1.3$\pm$1.4 &300.7 \\
192 &19.6$\pm$0.8$\pm$0.1 & 18.9 &301.4$\pm$3.3$\pm$1.5 & 291.4\\
196 &18.6$\pm$0.5$\pm$0.1 & 18.1 &285.8$\pm$2.0$\pm$1.5 & 279.6 \\
200 &18.2$\pm$0.5$\pm$0.1 & 17.4 &273.0$\pm$1.9$\pm$1.4 &268.5 \\
202 & 17.9$\pm$0.7$\pm$0.1& 17.0 &272.0$\pm$2.8$\pm$1.4 &263.3 \\
\end{tabular}
\end{ruledtabular}
\end{table}

\begin{table}
\caption{Upper limits on the $g_{L}^{2}/M_{L}^{2}$ for 
doubly charged scalar bileptons at 95\% C.L. Combination of 
results are also shown in the last row. Acollinearity 
cut $\theta_{acol}<10^{o}$ is considered and masses 
are in units of GeV. \label{tab2}}
\begin{ruledtabular}
\begin{tabular}{lcc}
$\sqrt{s}$ GeV & ${g_L^2}\over{m_L^2}$ & ${g_L^2}\over{m_L^2}$ \\
 &$|\cos\theta|< 0.7$ &$|\cos\theta|< 0.96$ \\
\hline
183 &$<5.5\times10^{-5}$   &$<4.1\times10^{-4}$ \\
189 &$<4.2\times10^{-5}$ &$<2.5\times10^{-4}$ \\
192 &$<5.5\times10^{-5}$ &$<3.2\times10^{-4}$ \\
196 &$<4.5\times10^{-5}$ &$<2.6\times10^{-4}$ \\
200 &$<4.3\times10^{-5}$ &$<2.4\times10^{-4}$ \\
202 &$<4.8\times10^{-5}$ &$<2.7\times10^{-4}$ \\
\hline
Combination& $<3.5\times10^{-5}$ &$<1.9\times10^{-4}$ \\
\end{tabular}
\end{ruledtabular}
\end{table}

\begin{table}
\caption{Upper limits on the $g_{L}^{2}/M_{L}^{2}$ for
doubly charged vector bileptons at 95\% C.L. Combination of
results are also shown in the last row. Acollinearity
cut $\theta_{acol}<10^{o}$ is considered. and masses
are in units of GeV. \label{tab3}}
\begin{ruledtabular}
\begin{tabular}{lcc}
$\sqrt{s}$ GeV & ${g_L^2}\over{m_L^2}$ & ${g_L^2}\over{m_L^2}$ \\
 &$|\cos\theta|< 0.7$ &$|\cos\theta|< 0.96$ \\
 \hline
 183 &$<2.0\times10^{-5}$   &$<1.8\times10^{-4}$ \\
 189 &$<1.3\times10^{-5}$ &$<1.0\times10^{-4}$ \\
 192 &$<2.0\times10^{-5}$ &$<1.4\times10^{-4}$ \\
 196 &$<1.5\times10^{-5}$ &$<1.1\times10^{-4}$ \\
 200 &$<1.4\times10^{-5}$ &$<1.0\times10^{-4}$ \\
 202 &$<1.7\times10^{-5}$ &$<1.2\times10^{-4}$ \\
 \hline
 Combination& $<9.8\times10^{-6}$ &$<7.4\times10^{-5}$ \\
 \end{tabular}
 \end{ruledtabular}
 \end{table}

\end{document}